\documentclass{iopart}

\usepackage{graphicx}
\usepackage{etoolbox}
\usepackage{iopams,wrapfig}
\usepackage{xspace}
\usepackage{subfig}
\usepackage{amsmath}
\usepackage[bottom]{footmisc}
\usepackage{hyperref}
\usepackage{graphicx}
\usepackage{cite}
\usepackage{csquotes}
\usepackage{float}
\usepackage{booktabs} 

\usepackage{xcolor} 
\usepackage{tabularx} 

\begin{document}

\title[]{Particle Identification with Deep Neural Networks Across Collision Energies in Simulated Proton–Proton Collisions}

\author{O. M. Khalaf$^{1}$, A. M. Hamed$^{1}$}

\address{$^{1}$Department of Physics, The American University in Cairo, New Cairo 11835, Egypt}

\begin{abstract}
This study demonstrates a proof-of-concept application of a deep neural network for particle identification in simulated high transverse momentum proton-proton collisions, with a focus on evaluating model performance under controlled conditions. A model trained on simulated Large Hadron Collider (LHC) proton-proton collisions at $\sqrt{s} = 13\,\mathrm{TeV}$ is used to classify nine particle species based on seven kinematic-level features. The model is then tested on simulated high transverse momentum Relativistic Heavy Ion Collider (RHIC) data at $\sqrt{s} = 200\,\mathrm{GeV}$ without any transfer learning, fine-tuning, or weight adjustment. It maintains accuracy above 91\% for both LHC and RHIC sets, while achieving above 96\% accuracy for all RHIC sets, including the $p_T > 7\,\mathrm{GeV}/c$ set, despite never being trained on any RHIC data. Analysis of per-class accuracy reveals how quantum chromodynamics (QCD) effects, such as leading particle effect and kinematic overlap at high $p_T$, shape the model's performance across particle types. These results suggest that the model captures physically meaningful features of high-energy collisions, rather than simply overfitting to kinematics of the training data. This study demonstrates the potential of simulation-trained deep neural networks to remain effective across lower energy regimes within a controlled environment, and motivates further investigation in realistic settings using detector-level features and more advanced network architectures. 
\end{abstract}

\textbf{keywords:} Deep learning, Particle identification, High transverse momentum, Proton-proton collisions, RHIC, LHC, Collider simulations
\newpage
\section{Introduction}
\noindent
Machine learning (ML), deep learning in particular, is playing an increasingly central role in data collection and analysis in modern particle physics especially at CERN's Large Hadron Collider (LHC) \cite{Radovic:2018dip}. Deep learning techniques have shown significant promise in particle identification tasks ranging from jet classification \cite{Guest:2018yhq}, trigger systems, to calorimetry and event reconstruction \cite{Belayneh2020}. Furthermore, such approaches have become critical in high momentum regimes where standard particle identification methods fail \cite{Acharya2021}. This large variety of machine learning applications has been primarily applied to the LHC data. In contrast, significantly fewer studies have explored similar techniques at the Relativistic Heavy Ion Collider (RHIC), despite its importance in studying the quark-gluon plasma (QGP) \cite{ADAMS2005102} and the proton's spin structure.\\

\noindent
Proton--proton collisions at RHIC, carried out at $\sqrt{s} = 200\,\mathrm{GeV}$ \cite{PhysRevLett.92.112301}, represent a very different kinematic regime compared to that of the LHC's $\sqrt{s} = 13\,\mathrm{TeV}$ \cite{Acharya2021,KHANDAI_2013}. These nuanced differences in kinematic distribution pose a meaningful challenge for evaluating the ability of deep learning models trained on one center-of-mass energy to perform when applied to another. While deep learning models show promising results in particle identification tasks, most ML efforts in high-energy physics are trained and tested on the same experimental or simulated data at the same energy scale \cite{Andrews2020,bhimji2017deepneuralnetworksphysics}.\\

\noindent
In this study, we evaluate the performance of a simple deep neural network trained exclusively on LHC proton-proton collisions to identify nine distinct but common particle types. The model was trained using seven kinematic-level features and tested on various $p_T > 3\,\mathrm{GeV}/c$ bins. Furthermore, we assessed the model's ability to remain effective at RHIC's lower center-of-mass energy, using identical high-$p_{T}$ selection criteria and event counts as the LHC set.\\

\noindent
To our knowledge, this is one of the earliest studies to explore the behavior of deep neural networks trained at LHC energies when applied to RHIC conditions in simulated proton-proton collisions. While neural networks have been used in HEP to interpolate over continuous variables like particle mass in simulated datasets \cite{Baldi2016}, those approaches were primarily focused on one energy regime, rather than evaluating performance across different collision energies. This manuscript further highlights both the potential and the limitations of simulation-trained models when applied across different collider center-of-mass energies. The analysis includes per-class accuracy and misclassification patterns, providing physics-motivated insight into the model's performance and limitations in the $p_T > 7\,\mathrm{GeV}/c$ regime across energies. Even though the study is limited to kinematics of simulated data, the results suggest that the model can learn meaningful physical features of proton-proton collisions. The findings motivate further investigation under realistic conditions, incorporating detector-level features and advanced network architectures, with some potential directions proposed in the conclusion.

\section{Dataset and Feature Design}
\noindent
PYTHIA 8\cite{sjostrand2015introduction} was used to simulate the p-p collision at both $\sqrt{s} = 13\,\mathrm{TeV}$ and $\sqrt{s} = 200\,\mathrm{GeV}$. The events were generated using PYTHIA version 8.3.07 with the Monash 2013 tune enabled by default. The default parton distribution function (PDF) used is NNPDF2.3 QCD+QED LO, which is consistent with the Monash tuning.\\

\noindent
Two constraints are applied to the generated data, a pseudorapidity cut $-2 < \eta < 2$, which corresponds to the central region of the detector, and a transverse momentum cut $p{_{_T}} > $ 3 GeV/c. The hadronic decay is enabled, which makes it necessary to choose particles that are long-lived and commonly produced in proton--proton collisions. This is because the model is not designed to deal with anomalies or particles that are quite rare and produced with low multiplicities. The particles selected are $\gamma$, $\pi^+$, $\pi^-$, $K^+$, $K^-$, $K_L^0$, $p$, $e^-$, $n$.\\

\noindent
 Particles like muons, hyperons, and various short-lived particles were not selected in order to maintain charge symmetry across particle types (3 positive, 3 neutral, 3 negative) and to focus on commonly produced particles with high multiplicities within the specified kinematic cuts. Moreover, the inclusion of relatively rare particles or those requiring specialized detectors and detection methods would introduce unnecessary complexity and class imbalance without contributing to the study's objective.\\
 
\noindent
The training set is a combination of two generated sets with slightly different parameters to facilitate data generation and event selection. It consists of a low $p_{T}$ set with only the pseudorapidity condition and a high $p_{T}$ set with a $p_T > 3\,\mathrm{GeV}/c$ cut. The training set contains 6.2 million events (5 million low  $p_{T}$ and 1.2 million high  $p_{T}$). Table~\ref{tab:event_cuts} shows the selected input features along with a brief description of them and Table~\ref{tab:pythia8_params} shows the generation parameters in detail for the training set.

\begin{table}[H]  
\caption{Summary of the Input Features and Their Physical Significance}
\label{tab:event_cuts}
\centering
\resizebox{\textwidth}{!}{ 
\begin{tabular}{c c c}
\hline\hline   
\textbf{Input Feature} & \textbf{Physical Significance} &  \textbf{Equation}\\ \hline
Energy & Energy from the relativistic energy-momentum relation & $E^2 = |\vec{p}|^2 + m^2$ \\
Momentum & Magnitude of the momentum vector & $|\vec{p}| = \sqrt{p_x^2 + p_y^2 + p_z^2}$ \\
Transverse Momentum & Magnitude of momentum in the transverse plane & $p_T = \sqrt{p_x^2 + p_y^2}$ \\
Pseudorapidity & Spatial angle relative to the beam axis & $\eta = -\ln\left[\tan\left(\frac{\theta}{2}\right)\right]$ \\
Rapidity & Lorentz-invariant measure along the beam direction & $y = \frac{1}{2} \ln\left(\frac{E + p_z}{E - p_z}\right)$ \\
Azimuthal Angle & Angle in the transverse plane around beam axis & $\phi = \arctan\left(\frac{p_y}{p_x}\right)$ \\
Charge & Electric charge of the particle in units of $e$ & $q$ \\ \hline
\end{tabular}
} 
\end{table}

\begin{table}[htbp!]  
\centering
\caption{Summary of Key PYTHIA8 Parameters Used in the Training Set}
\label{tab:pythia8_params}
\centering 
\begin{tabular}{c c c} 
\hline\hline   
\textbf{Parameter Name} & \textbf{Low $p_T$} & \textbf{High $p_T$}  \\ \hline
\texttt{Beams:idA}           & 2212 (proton)      & 2212 (proton) \\
\texttt{Beams:idB}           & 2212 (proton)      & 2212 (proton) \\ 
\texttt{Beams:eCM}           &   13 TeV &   13 TeV\\ 
\texttt{Number of Events}    & 5 million  & 1.2 million \\ 
\texttt{PhaseSpace:pTHatMin} &  0 GeV    & 2.0 GeV \\ 
\texttt{HardQCD:all}         & \texttt{on}  & \texttt{on} \\
\texttt{Charmonium:all}      & \texttt{off} & \texttt{on} \\ 
\texttt{Bottomonium:all}     & \texttt{off}   & \texttt{on}\\ 
\texttt{PromptPhoton:all}    & \texttt{on}  & \texttt{on}\\
\texttt{Random:setSeed}      & \texttt{on}   & \texttt{on} \\ 
\texttt{PartonLevel:ISR}     & \texttt{on}  & \texttt{on} \\ 
\texttt{PartonLevel:FSR}     & \texttt{on} & \texttt{on}\\ 
\texttt{HadronLevel:Decay}   & \texttt{on} & \texttt{on}\\ \hline
\end{tabular}
\end{table}

\section{Kinematic Distributions and Particle Composition}
\begin{figure}[h!]
\centering
\begin{tabular}{ccc}
\includegraphics[width=0.3\textwidth]{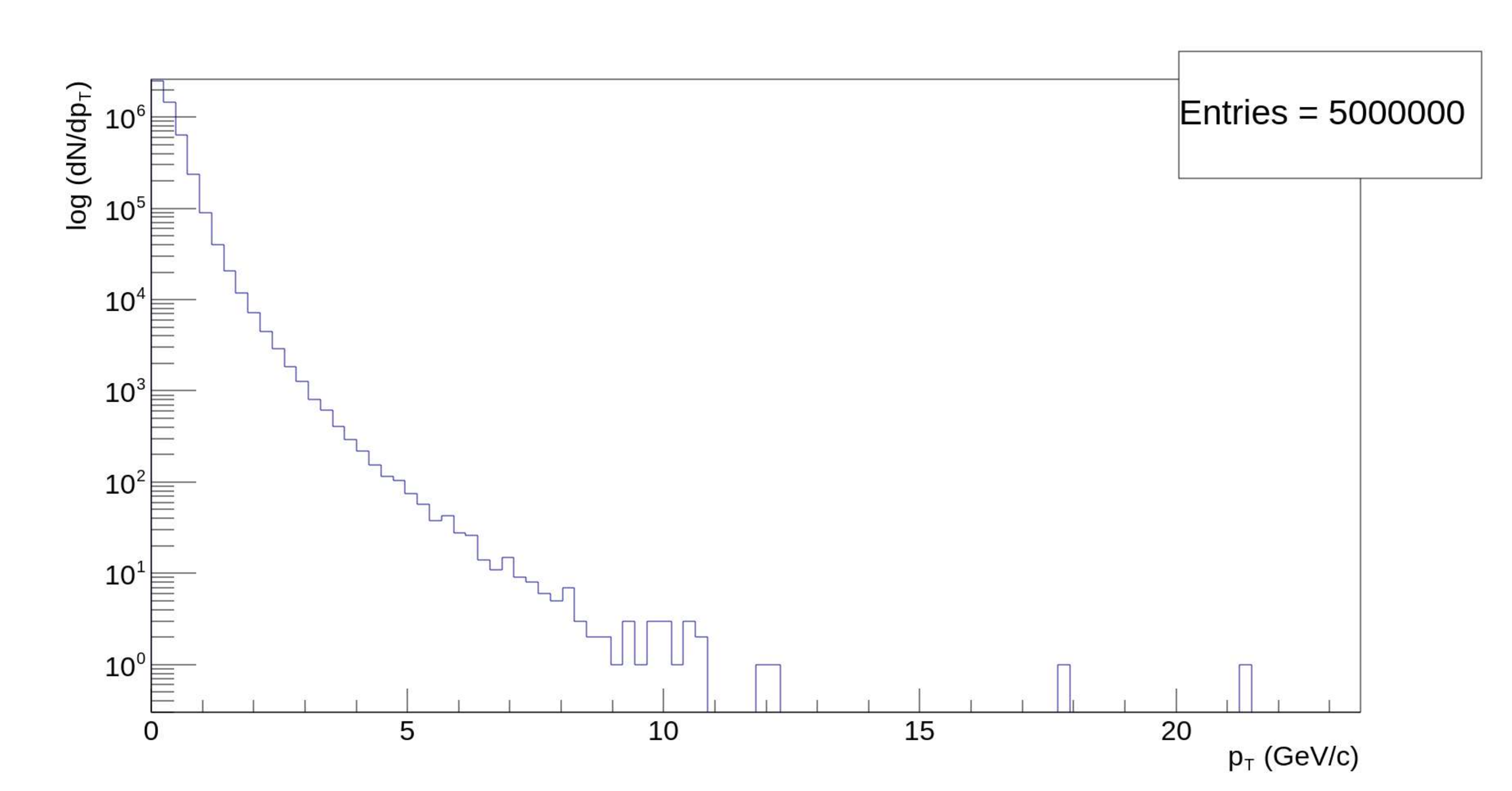} &
\includegraphics[width=0.3\textwidth]{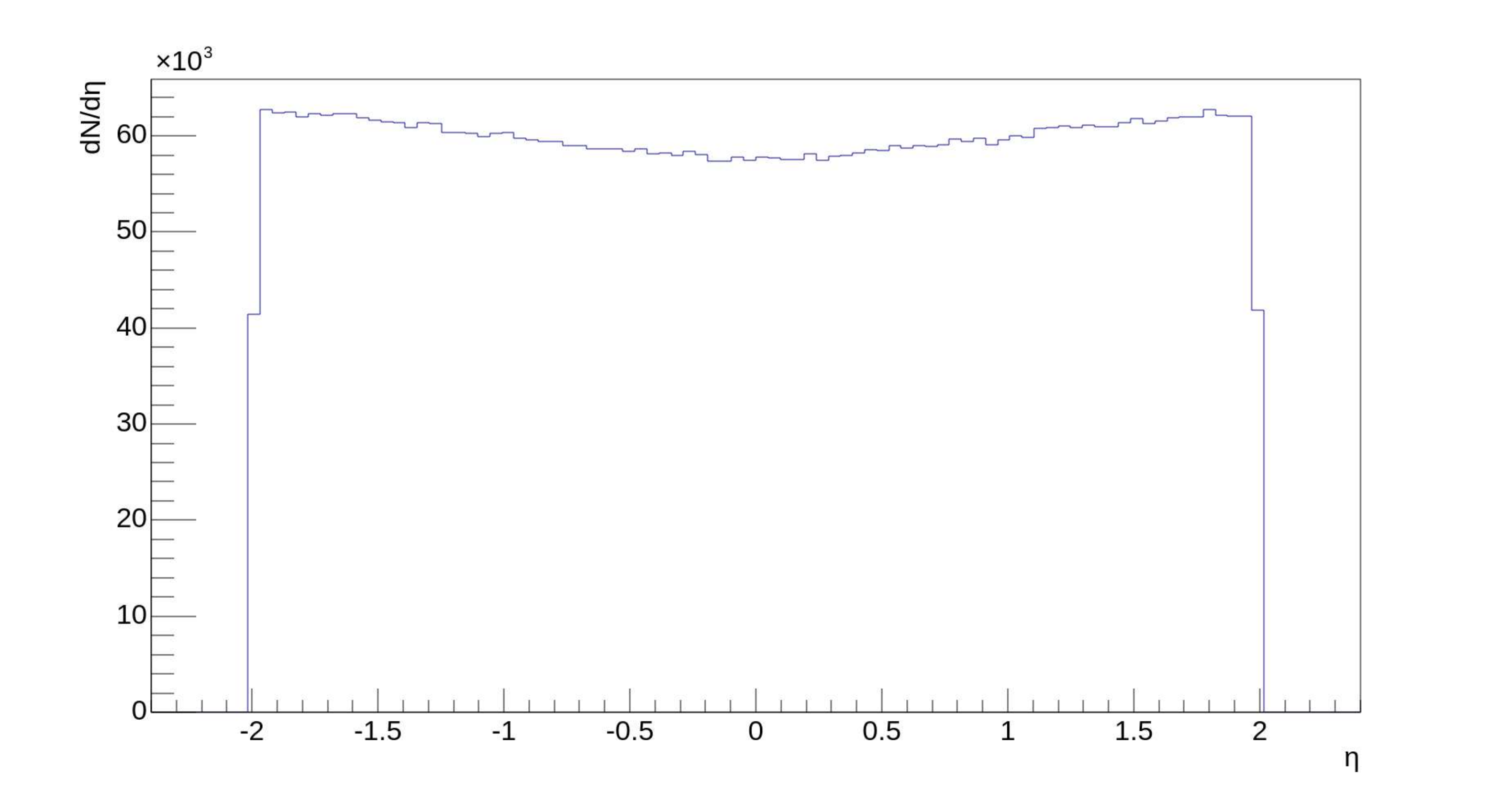} &
\includegraphics[width=0.3\textwidth]{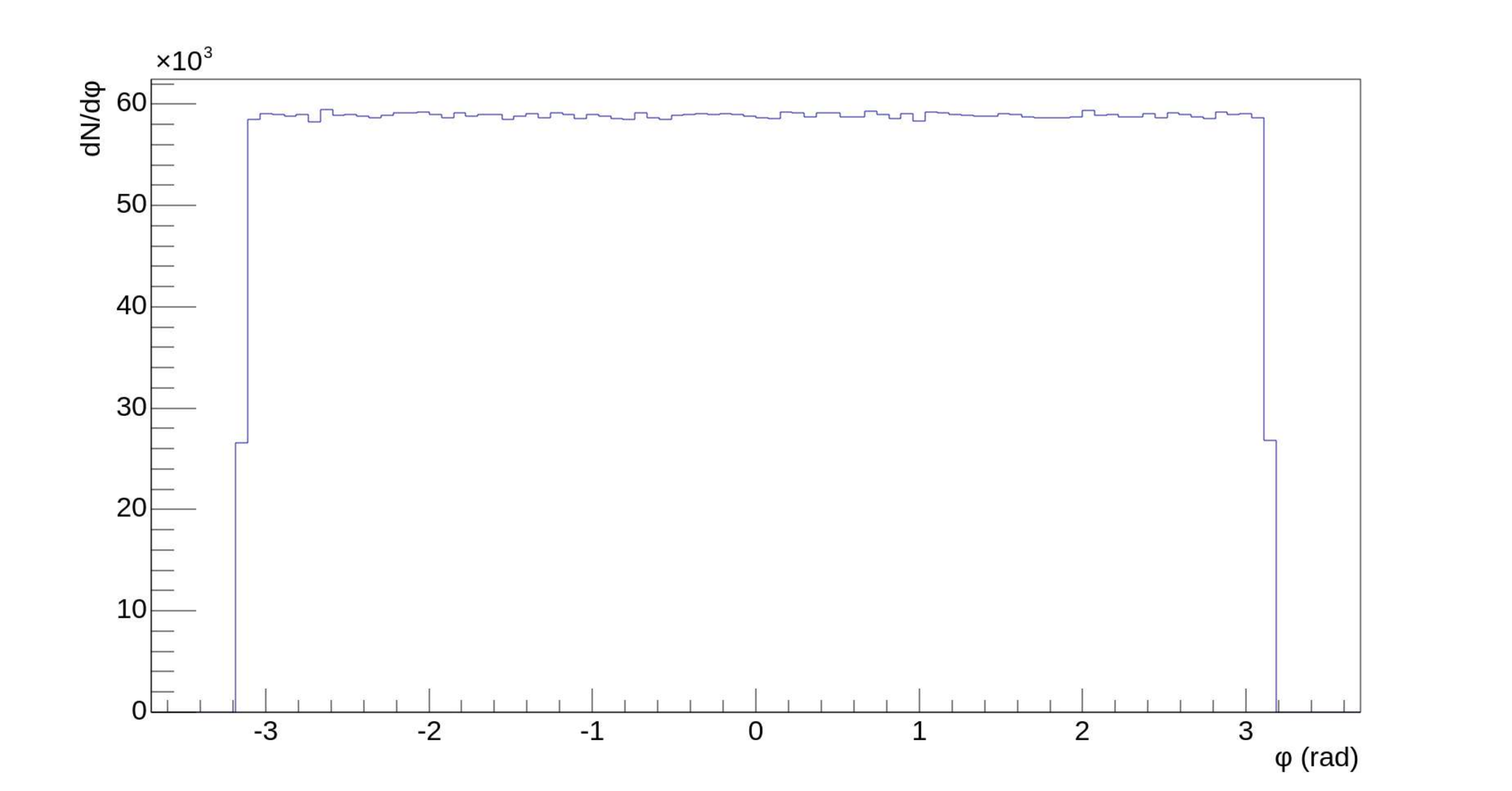} \\
\includegraphics[width=0.3\textwidth]{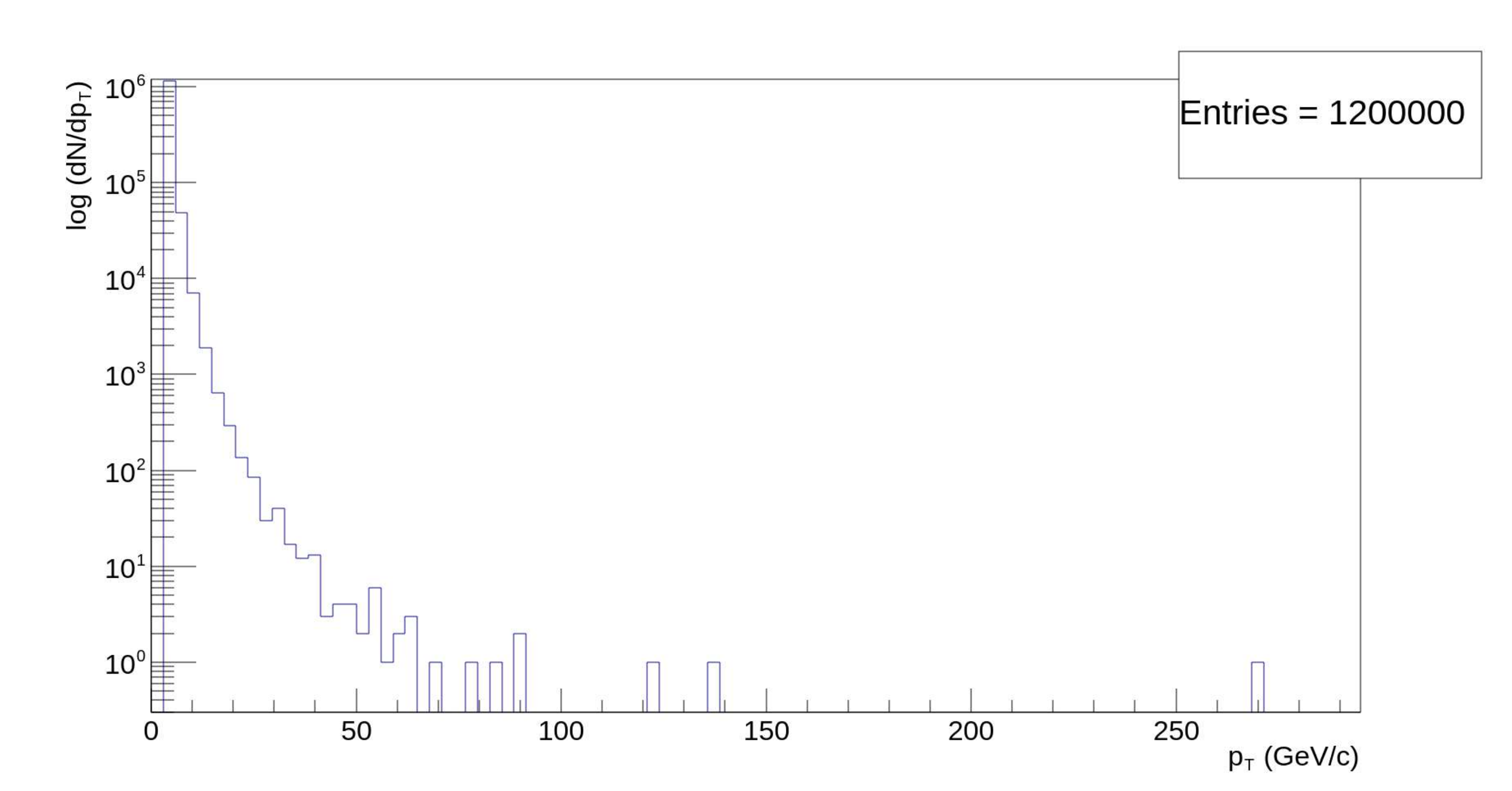} &
\includegraphics[width=0.3\textwidth]{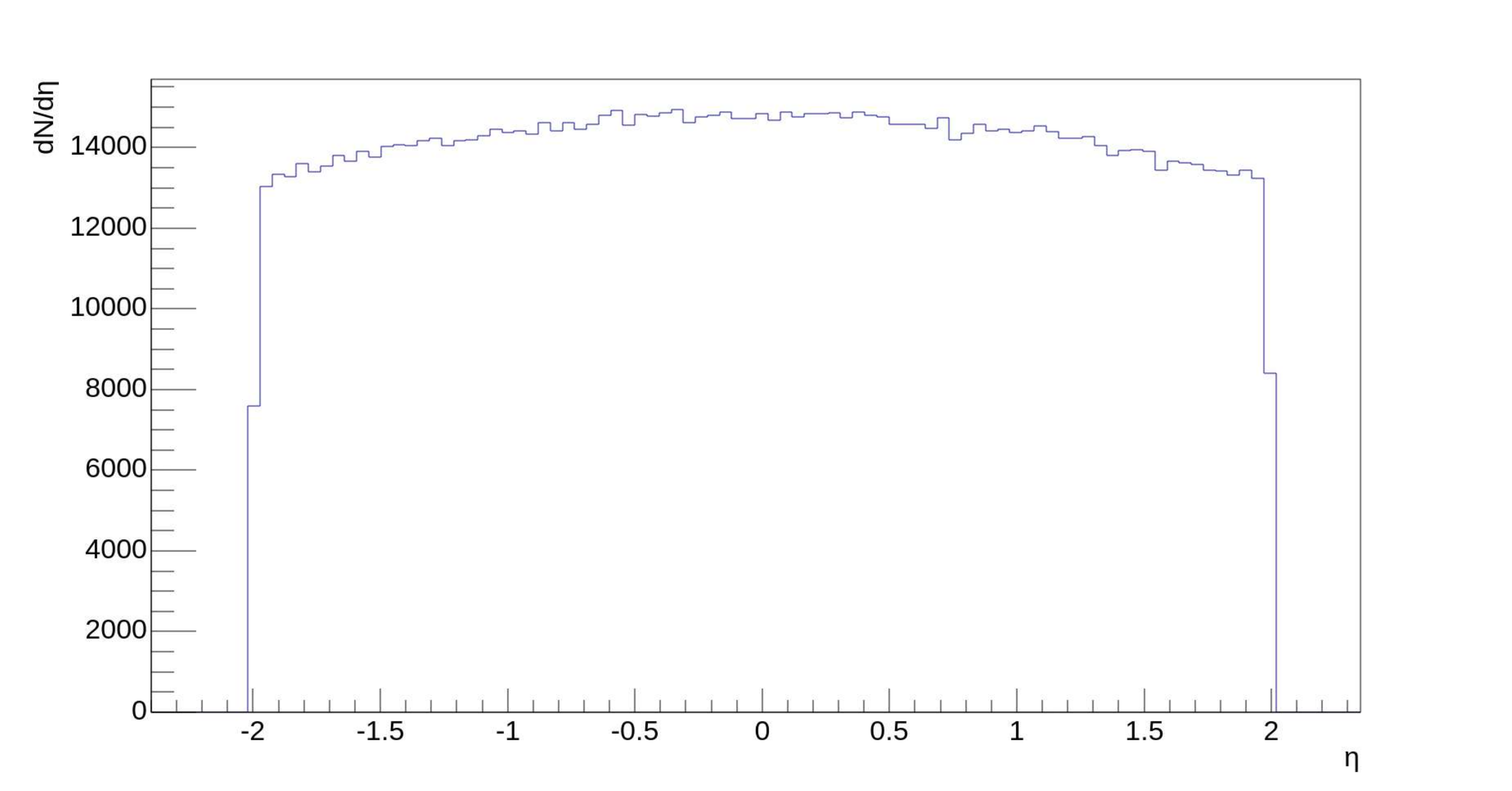} &
\includegraphics[width=0.3\textwidth]{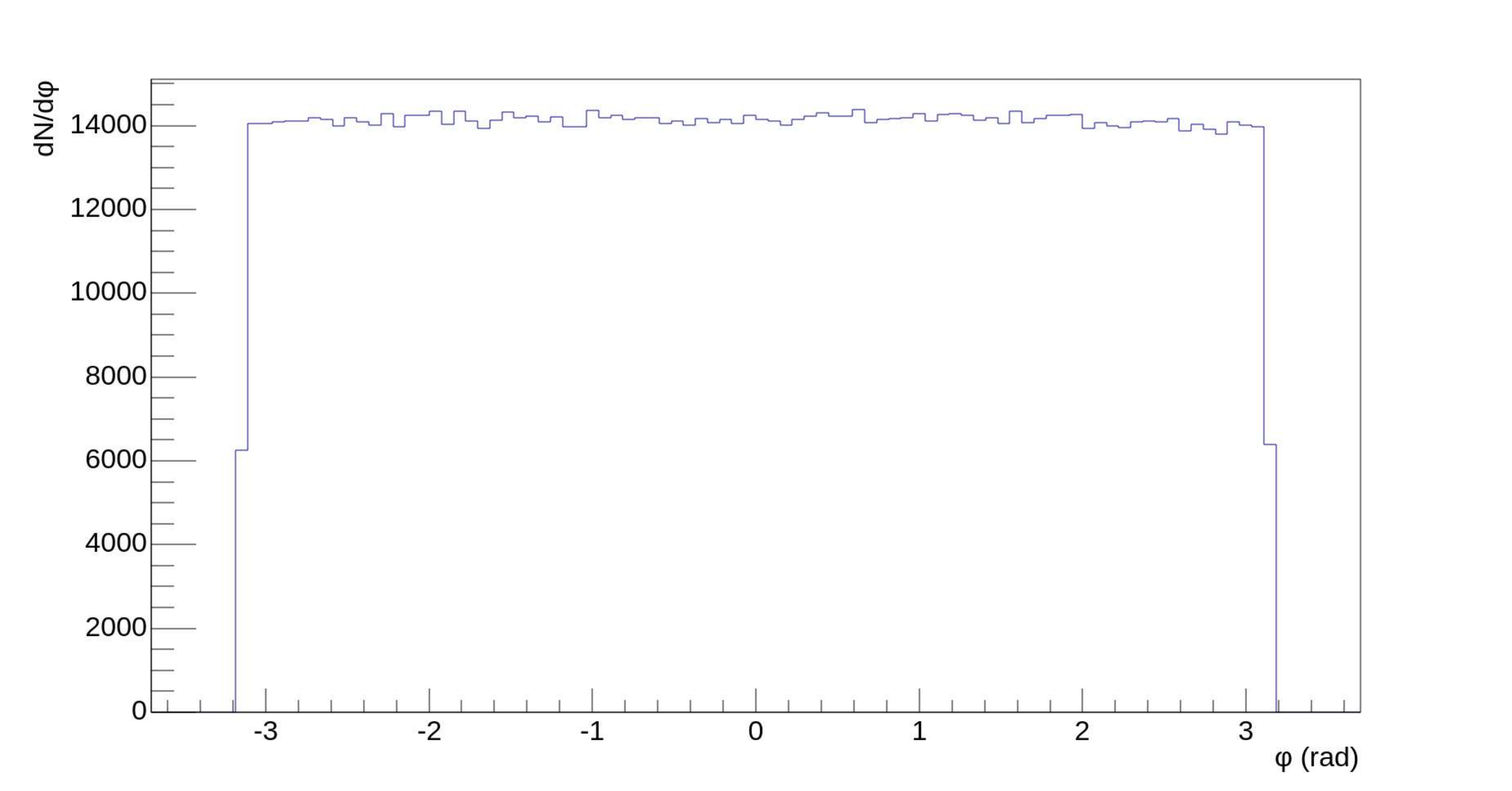} \\
\end{tabular}
\caption{Summary of $p_T$, $\eta$, and $\phi$ distributions for low-$p_T$ and high-$p_T$ sets. The top row corresponds to the low-$p_T$ set and the bottom row corresponds to the high-$p_T$ set.}
\label{fig:features}
\end{figure}

\noindent
The input parameters are selected based on the criterion that they should be obtainable from laboratory measurements or derivable from other measured quantities. Moreover, the detectors that inspire the feature selection should be available at both the LHC and RHIC. Based on that information, the detectors responsible for the measurements would be hadronic and electromagnetic calorimeters\cite{BROWN201247,Fabjan:2003aq}, time projection chambers\cite{YU201355,ANDERSON2003659}, and silicon vertex detectors \cite{QIU20141141,Buckland:2022ism}. These detectors are foundational and essential for particle colliders. There is an additional constraint on the parameters, which is that they must be able to be generated using the event generator \texttt{Pythia~8}. The distributions of the key input features for both the low- and high-$p_T$ sets are shown in Figure~\ref{fig:features}, illustrating the kinematic differences between the two datasets.\\

\begin{table}[H]
\caption{Charge and particle distribution percentages in low and high transverse momentum training sets.}
\label{tab:particle_distribution}
\centering
\begin{tabular}{c c c c}
\hline\hline
\textbf{Particle} & \textbf{Charge} & \textbf{\% in low $p_T$} & \textbf{\% in high $p_T$} \\
\hline
$\gamma$ & 0 & 52.193 & 11.198 \\
$K_L^0$ & 0 & 1.791 & 8.766 \\
$n$ & 0 & 1.016 & 6.578 \\
$\pi^+$ & +1 & 20.045 & 24.708 \\
$p$ & +1 & 1.024 & 6.580 \\
$K^+$ & +1 & 1.812 & 8.714 \\
$e^-$ & $-1$ & 0.328 & 0.107 \\
$K^-$ & $-1$ & 1.806 & 8.661 \\
$\pi^-$ & $-1$ & 19.983 & 24.688 \\
\hline
\end{tabular}
\end{table}

\subsection{Momentum and Angular Distributions}
For the low $p_T$ regime, the transverse momentum distribution exhibits a steep falloff consistent with a power-law spectrum at high $p_T$\cite{2cc630d9da064d4f92141b7cb8429fc9}. This reflects the dominance of soft QCD processes. Most particles in this regime fall within the $0$--$2\,\mathrm{GeV}/c$ region. Moreover, only a small fraction of particles have higher $p_T$ due to the scarcity of hard scattering processes. This distribution is qualitatively similar to the full momentum distribution, as the transverse momentum represents a component of the total momentum. The logarithmic scale was chosen to highlight the tail of the curve and the large dynamic range of particle transverse momentum values.\\

\noindent
The pseudorapidity distribution is symmetric around zero, as expected for particles produced in the central rapidity region in symmetric proton--proton collisions\cite{CMS:2015zrm}. The $-2 < \eta < 2$ cut corresponds to a polar angle between $164.5^\circ$ and $15.5^\circ$, avoiding regions too close to the beamline. This distribution closely matches the rapidity distribution at high momentum under the condition that $p \gg m$. The azimuthal distribution is completely symmetric and uniform, indicating isotropic particle production. Moreover, the values of the azimuthal angle span the full range $[-\pi, \pi]$, providing full coverage around the beam axis.

\subsection{Low-$p_T$ Particle Composition}
In the low $p_T$ regime shown in Table~\ref{tab:particle_distribution}, the most dominant particles are neutral particles, which is natural since approximately $52\%$ of the particles are photons. These photons are primarily produced from the decay of neutral pions via $\pi^0 \rightarrow 2\gamma$. This decay has a branching ratio of approximately $98.8\%$ \cite{ParticleDataGroup:2022pth}, and is a major contributor to the photon abundance. Additionally, photons can be produced directly during collisions or as byproducts of other decays.\\

\noindent
The second most abundant particles are the positive and negative pions, each constituting about $20\%$ of the particles. Their abundance is due to their low mass, making them kinematically favored in QCD interactions. Their numbers are equal due to isospin symmetry, as $\pi^+$, $\pi^0$, and $\pi^-$ form an isospin triplet \cite{Griffiths2008}. Had the neutral pion not decayed promptly, the number of $\pi^0$ would have matched the $\pi^\pm$ counts, but its rapid decay into two photons effectively boosts the photon yield relative to charged pions.\\

\noindent
Protons and neutrons, which form an isospin doublet \cite{Griffiths2008}, are produced in approximately equal numbers due to having similar mass and quark structure. Their overall lower production rate compared to pions is attributed to their higher mass, making their production less kinematically favorable.\\

\noindent
Kaons, which include a strange quark or antiquark, are produced in approximately equal numbers between $K^+$ and $K^-$ as a reflection of strangeness conservation. However, strange quarks are subject to strangeness suppression \cite{Drescher_2002}, leading to fewer kaons compared to pions. The kaon-to-pion ratio ($K/\pi$) calculated as $0.135$ is consistent with typical values observed in high-energy proton-proton collisions \cite{CMS:2017eoq}.\\

\noindent
Finally, electrons represent the least abundant particle type because they are produced via electromagnetic processes, which are less dominant compared to strong interactions in these collisions.

\subsection{High-$p_T$ Behavior and Transitions}
Compared to the low-$p_T$ regime, the high-$p_T$ momentum distribution decreases more gradually, with the falloff occurring around $50\,\mathrm{GeV}/c$ compared to the previous $10\,\mathrm{GeV}/c$. Most particles in the high-$p_T$ sample have momenta between $3$--$5\,\mathrm{GeV}/c$, with rare high energy particles reaching up to $100$--$200\,\mathrm{GeV}/c$ in some cases. The pseudorapidity distribution remains completely uniform and symmetric, showing no drop at $\eta=0$. The azimuthal distribution retains its uniformity, consistent with the low-$p_T$ data.\\

\noindent
The particle type and charge distributions exhibit notable shifts at high transverse momentum. Photons, previously dominant at low $p_T$, now account for only about $11\%$ of the total. While the $\pi^0 \rightarrow 2\gamma$ decay process still occurs, the resulting photons often fall below the $p_T > 3\,\mathrm{GeV}/c$ threshold. Consequently, the charged pion contribution rises, with $\pi^+$ and $\pi^-$ together making up about $50\%$ of the particles.\\

\noindent
The kaons experience a significant boost in their production, rising from approximately $1.8\%$ to $8.7\%$ for both $K^+$ and $K^-$. This increase is attributed to the greater availability of energy in collisions, partially overcoming strangeness suppression. Similarly, protons and neutrons also exhibit increased production due to the dominance of hard scattering and fragmentation processes at higher energies. Even at high $p_T$, electrons still constitute only about $0.1\%$ of all produced particles, again highlighting the influence of strong interaction processes.
\subsection{Suitability for Model Training}
All previously examined kinematic features and distributions are consistent with theoretical expectations and align well with experimental observations from proton-proton collisions. The transition from a photon-dominated distribution to a more dynamic and energetic particle composition reflects the increasing influence of hard scattering and fragmentation processes. The $|\eta| < 2$ cut focuses the analysis on the central detector region, where particle identification is most effective. Combining soft particles from the low momentum set with rarer, high-momentum particles ensures coverage of a broad range of the kinematic spectrum of proton-proton collisions. The physical validity and diversity of the combined dataset make it highly suitable for training the model to perform particle identification across a wide momentum range.

\section {Neural Network Training}
The neural network architecture chosen is a standard feed-forward network, specifically a multi-layer perceptron (MLP)\cite{Rumelhart1986}. The hidden layers use the Rectified Linear Unit (ReLU) activation function, a computationally efficient nonlinear activation function defined as\cite{kunc2024decadesactivationscomprehensivesurvey}
\begin{equation}
f(x) = \max(0, x).
\end{equation}
Compared to other alternatives such as tanh or sigmoid, ReLU avoids  exponentiation. This makes it suitable for handling large datasets such as those from LHC simulations. Non-linearity in hidden layers is essential in capturing complex relationships between inputs and outputs.\\

\noindent
The output layer activation function and the loss function are closely related and therefore are discussed together. The Softmax activation function is well suited for multi-class classification which is the goal of the network. The function converts the values it receives from the hidden layer into a probability distribution of the classes, with the sum of probabilities being equal to one. The Softmax function takes the form: \cite{nwankpa2018activationfunctionscomparisontrends}

\begin{equation}
f(x_i) = \frac{e^{x_i}}{\sum_{j} e^{x_j}}
\end{equation}
\noindent
where $f(x_i)$ is the predicted probability, the denominator normalizes the output so that the sum of probabilities is equal to one. This predicted class corresponds to  the highest output probability.\\

\noindent
The network is trained to minimize the categorical cross-entropy loss, defined as \cite{Kerkhof2023}
\begin{equation}
\text{CCE} = -\frac{1}{N} \sum_{i=1}^{N} \sum_{j=1}^{c} y_{ij} \log\left(\hat{y}_{ij}\right)
\end{equation}
where $y_{ij}$ represents the true label of the class while $\hat{y}_{ij}$ represents the predicted probability of the class (output of the Softmax function). $N$ is the total number of samples and $c$ is the number of classes. This function measures how close the predicted distribution is to the true distribution of class labels.\\

\noindent
The optimizer is responsible for updating the weights of the model during training by adjusting key parameters like the learning rate. For this current setup, the Adaptive Moment Estimation (Adam) optimizer is used \cite{kingma2017adammethodstochasticoptimization}. It is adaptive as it adjusts the individual learning rates of each parameter. It combines the benefits of Root Mean Square Propagation(RMSProp)\cite{ruder2017overviewgradientdescentoptimization}, including adaptive learning rates, with additional advantages such as bias correction and momentum, making it well-suited  for the network at hand.\\

\noindent
For the network’s structure, each input parameter is assigned to one neuron in the input layer; therefore, the input layer has seven neurons. The number of hidden layers depends on the complexity of the task, but having two or more makes the network a deep neural network. Deep neural networks are suitable for tasks where standard methods struggle, due to their ability to capture complex patterns. The number of neurons in the hidden layer has a wide range of integer values depending on the application. However, a good starting point is having an equal number of neurons as the input layer. This network underwent many different iterations before settling on the final structure. The initial structure was 7 neurons in both hidden layers; however, to reduce overfitting, the number of neurons was reduced to 5 in the second hidden layer. Finally, the output layer uses a Softmax activation function, where each neuron represents the probability of a class. Since there are nine particle types, the output layer contains nine neurons.\\

\begin{figure}[H]
\centering
\includegraphics[width=100mm]{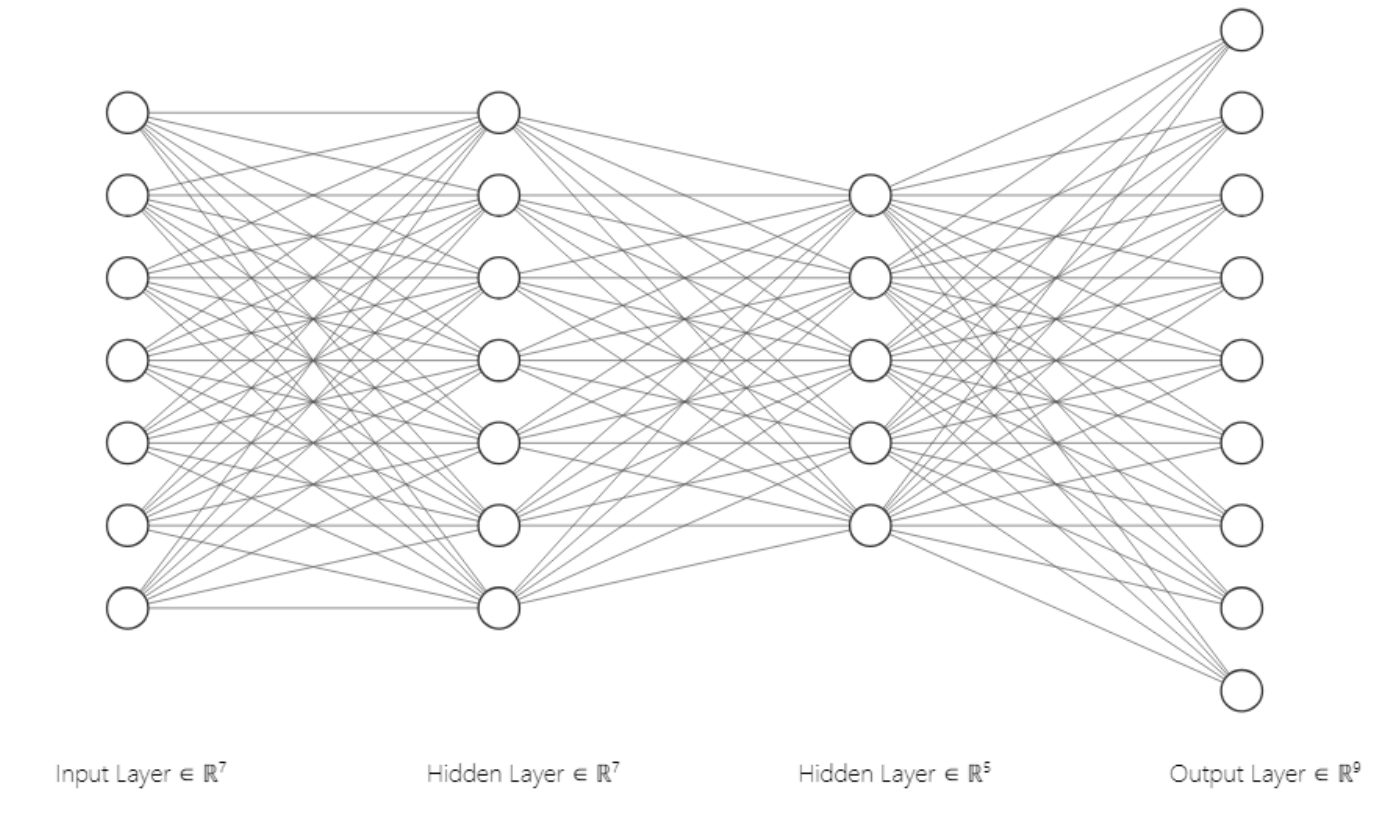}
\caption{ Structure of the deep neural network}
\label{fig:Structure}
\end{figure}

\noindent
The network was trained for 34 epochs with a batch size of 128, taking an average of 41.5 seconds per epoch and about 23.5 minutes to train. The chosen regularization method is early stopping\cite{Prechelt2012} with a patience value of 8. The model continues iterating until there are no improvements in validation loss for 8 consecutive epochs. Afterwards, the model restores the weights from the epoch with the lowest validation loss. This training technique improves generalization and reduces overfitting on future test sets.\\

\noindent
The dataset is split into three parts: training, validation, and test sets. The training set is used by the model to learn the relationships and patterns between inputs and outputs. The model iterates over the training data for 34 epochs. During training, the model receives the inputs and the true label or identification of the particle to learn the mapping between the input parameters and the output labels.\\

\noindent
The validation consists of examples which the model hasn’t seen. It is used after each epoch to check overall how it generalizes to unseen data. The validation set is crucial in guiding and improving the performance of the model and avoiding issues such as overfitting, where the model performs well in training but fails to generalize to new examples.\\

\noindent
The test set is used to assess the overall performance of the model after applying all the improvements and tuning during training. It is only used once, after the model has finished training, to examine how the model performs on real unseen data.\\

\begin{figure}[H]
\centering
\includegraphics[width=150mm]{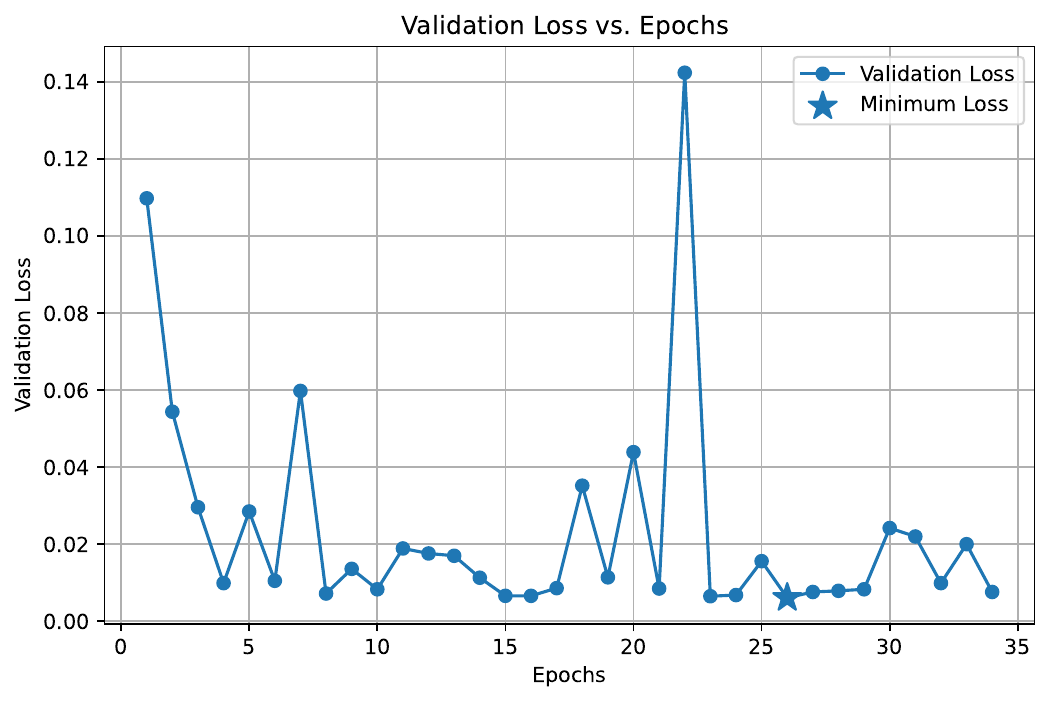}
\caption{ Validation loss vs. epochs}
\label{fig:validation_loss}
\end{figure}

\noindent
The dataset is split into an 80\% training, 10\% validation, and 10\% test sets. Although the 90-5-5 split is common, the 80-10-10 split was chosen to maintain a large number of examples, in this case about 5 million training examples while still having a considerable validation and test set of 620 thousand examples. Figure~\ref{fig:validation_loss} shows the change in validation loss during training. The point of minimum loss is highlighted by a star.\\

\noindent
The normalization of the training data was considered but it was not applied. By observing the values of all the input parameters, they all fall within manageable ranges where the minimum is $-3.14$ (from the azimuthal angle). In terms of transverse momentum, most particles fall within a range of $0$ to $50\,\mathrm{GeV}/c$, with a small number of outliers having $100$ to $200\,\mathrm{GeV}/c$. For the low momentum set without the $p_T > 3\,\mathrm{GeV}/c$ cut, most of the generated particles have a transverse momentum value between $0$ and $2\,\mathrm{GeV}/c$. Since the values for all input parameters are within close range of each other, normalization was not expected to affect the model's performance significantly. The robust and consistently high accuracy performance of the model across different seeds in Table~\ref{tab:training_summary}, is an indication that normalization was not a limiting factor for performance in this study. \\

\begin{table}[ht!]
\caption{Summary of the training, validation, and test set performances of the neural network.}
\label{tab:training_summary}
\centering
\begin{tabular}{c c c c}
\hline\hline
\textbf{Point of Comparison} & \textbf{Training Set} & \textbf{Validation Set} & \textbf{Test Set} \\
\hline
Accuracy (\%) & 99.47 & 99.88 & 99.88\,\textpm\,0.035 \\
Loss & 0.0172 & 0.0061 & 0.0058 \\
Number of examples & 4,960,000 & 620,000 & 620,000 \\
\hline
\end{tabular}
\end{table}

\begin{figure}[H]
\centering
\includegraphics[width=100mm]{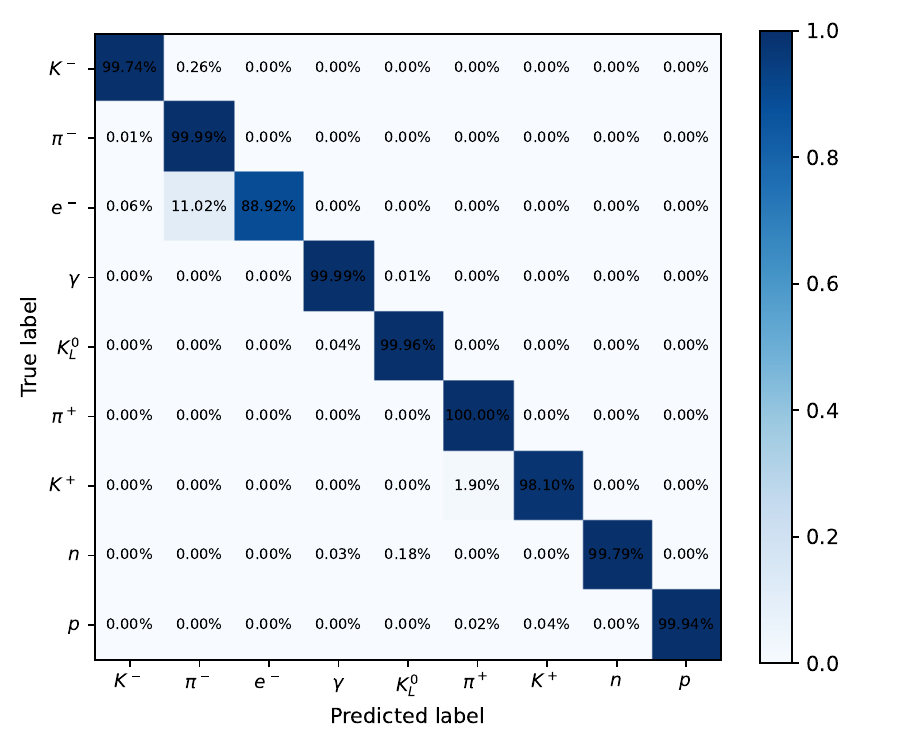}
\caption{ Confusion matrix for the test set (split from the $13\,\mathrm{TeV}$ LHC data), normalized row wise. The diagonal represents the correctly identified particles while the non diagonal elements represent misidentified particles.}
\label{fig:confusion_test}
\end{figure}

\noindent
Accuracy is defined as the ratio of correctly classified particles to the total number of particles in a given set:
\begin{equation}
    \text{Accuracy} = \frac{\text{Number of correctly classified particles (all classes)}}{\text{Total number of particles (all classes)}}
\end{equation}
 As shown in Table~\ref{tab:training_summary}, the model performs well on all three sets, having comparable accuracy above 99\%. The lowest accuracy observed is the training set, which can be attributed to its large size while the validation and test sets represent only a small subset of the total data. This means the training set contains a broader range of high momentum events, which explains its slightly reduced accuracy.\\

\noindent
To evaluate the robustness of the model, five training runs were performed with different seeds, which correspond to different weight initializations and optimizer behavior during training. For the full-range LHC test set, the resulting accuracies varied between 99.80\% and  99.88\%, with a mean value of 99.86\% and a sample standard deviation of 0.035. This small variation indicates that the model generalizes well to unseen data, and that its performance is stable and independent of initialization and stochastic training effects.\\

\noindent
Figure~\ref{fig:confusion_test} shows the identification accuracy for each particle type, as well as the misidentification pattern for each class. The per-class accuracy is defined as the ratio of correctly classified particles of a given type to the total number of actual particles of that type in the set:
\begin{equation}
\text{Per-class accuracy} = \frac{\text{Number of correctly classified particles of type } X}{\text{Total number of particles of type } X}
\end{equation}
This definition is equivalent to efficiency in particle identification. While purity, the fraction of correctly predicted particles of a given type relative to the total number of predicted particles of the same type, is a common metric, we focus primarily on evaluating model generalization across collision energies. For this reason, we emphasize overall classification accuracy across the two center-of-mass energies. However, purity values can be obtained directly from confusion matrices if needed.\\

\noindent
The average per-class accuracy is 98.5\%, with electrons exhibiting the lowest per-class classification accuracy. This reduced accuracy is mainly attributed to the kinematic similarity between electrons and negative pions, leading to the frequent misidentification of electrons as $\pi^-$. Moreover, electrons are the least represented class in the dataset, making it more difficult for the model to learn and distinguish their kinematic features from those of a more dominant class, such as pions.\\

\noindent
Overall, the model has no issues generalizing to unseen data based on the high performance and low loss values for both validation and test sets. The model is subsequently tested on specific cuts of high transverse momentum subsets at both RHIC and LHC energies.

\section{Test Sets Results}
The model was tested on both RHIC and LHC sets at a transverse momentum greater than $3\,\mathrm{GeV}/c$. The high-$p_T$ range was probed further by dividing them into $p_T$ bins with $3$-$5$, $5$-$7$, and $>7\,\mathrm{GeV}/c$. The data were generated with the same conditions as the high-$p_T$ of the training data in Table~\ref{tab:pythia8_params}. The only change was setting \texttt{pTHatMin}=4.0 GeV for convenience and to facilitate event generation. The model was tested specifically on RHIC to verify its ability to generalize to lower center-of-mass energy, as it was trained on LHC’s $\sqrt{s} = 13\,\mathrm{TeV}$. The performance results are summarized in Table~\ref{tab:test_set_results}.\\

\begin{table}[H]
\caption{Summary of the model's performance on the RHIC and LHC test sets across different $p_T$ ranges.}
\label{tab:test_set_results}
\centering
\resizebox{\textwidth}{!}{ 
\begin{tabular}{c c c c c c c c c}
\hline\hline
\textbf{Point of Comparison} & \textbf{High $p_T$ LHC} & \textbf{High $p_T$ RHIC} & \textbf{3-5 LHC} & \textbf{3-5 RHIC} & \textbf{5-7 LHC} & \textbf{5-7 RHIC} & \textbf{7+ LHC} & \textbf{7+ RHIC} \\
\hline
Accuracy (\%) & 99.44 & 99.70 & 99.80 & 99.87 & 98.24 & 98.66 & 91.33 & 96.23 \\
Loss & 0.0291 & 0.0180 & 0.0181 & 0.0129 & 0.0716 & 0.0467 & 0.2380 & 0.1108 \\
Number of examples & 500,000 & 500,000 & 50,000 & 50,000 & 50,000 & 50,000 & 50,000 & 50,000 \\
\hline
\end{tabular}
} 
\end{table}

\noindent
The model achieves high performance in the high transverse momentum regions, retaining an accuracy above 91\% for all test sets at both LHC and RHIC. Moreover, it performs as expected, having slightly lower accuracy as it moves into higher transverse momentum regions.

\section{Discussion}
\begin{figure}[H]
\centering
\begin{tabular}{cc}
\includegraphics[width=0.45\textwidth]{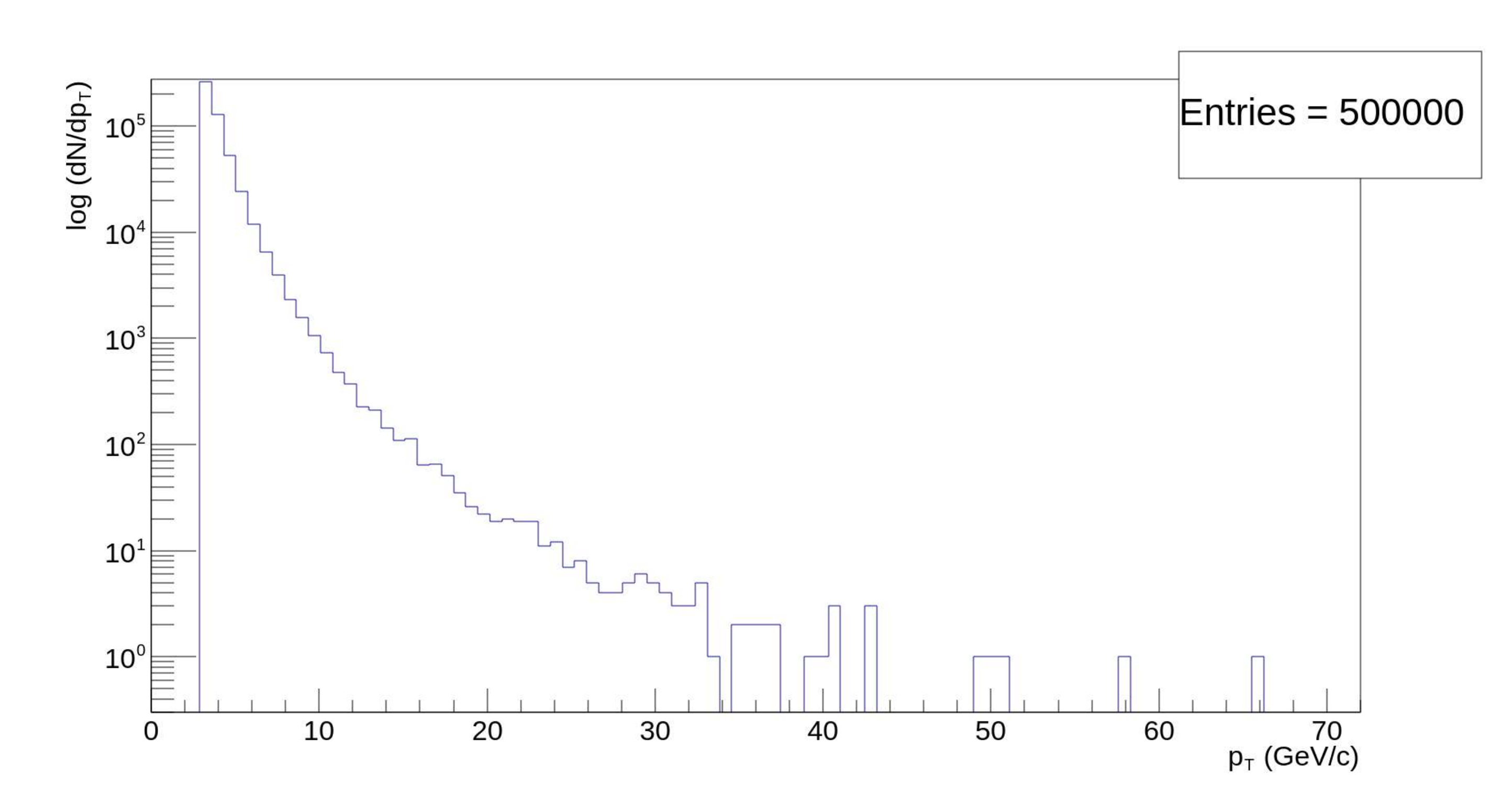} &
\includegraphics[width=0.45\textwidth]{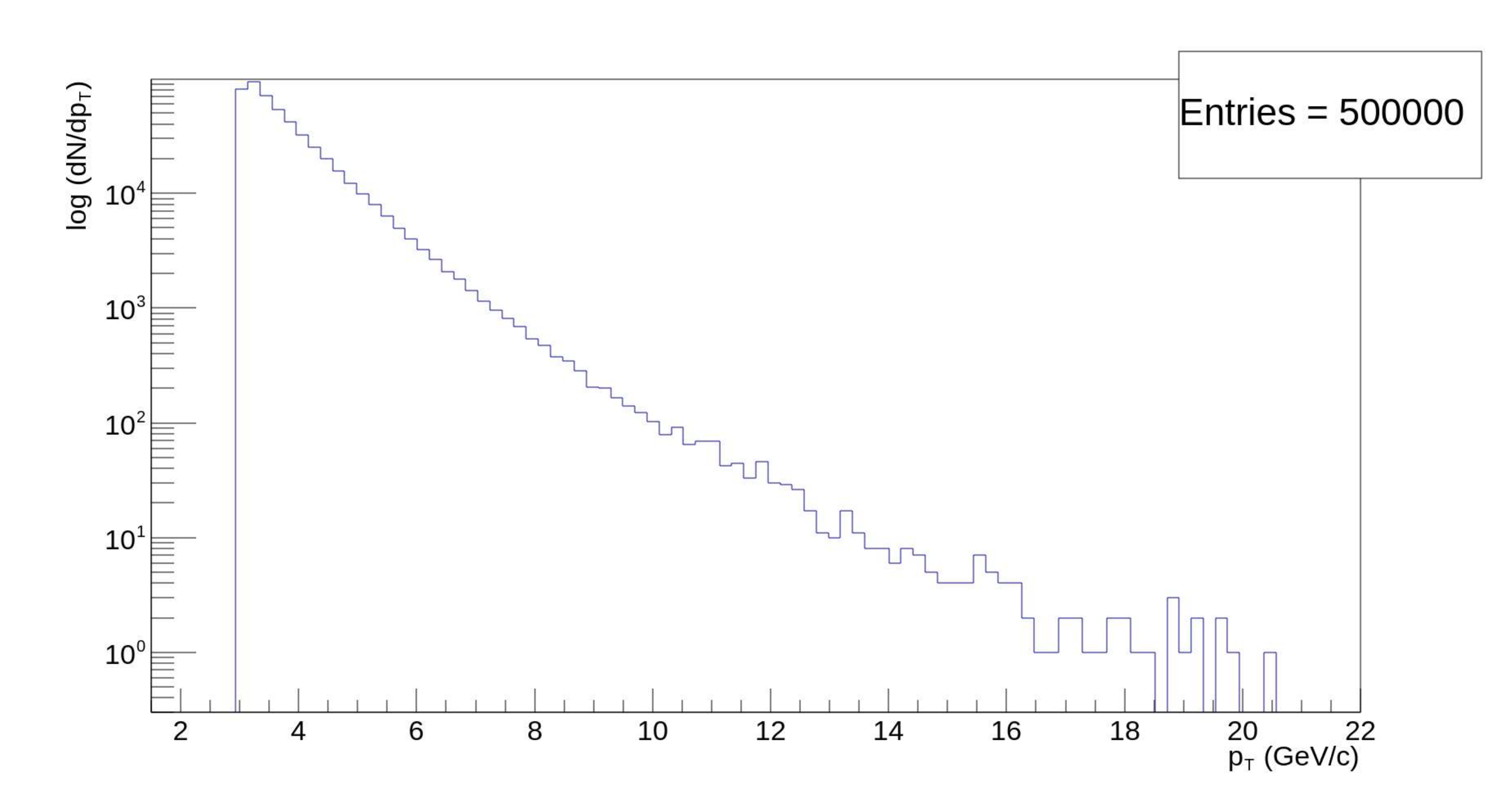} \\
\textbf{(a)} LHC $p_T$ distribution & \textbf{(b)} RHIC $p_T$ distribution \\
\end{tabular}
\caption{Comparison of transverse momentum ($p_T$) distributions between (a) LHC $\sqrt{s} = 13\,\mathrm{TeV}$ proton-proton collisions and (b) RHIC $\sqrt{s} = 200\,\mathrm{GeV}$ proton-proton collisions with $p_T > 3\,\mathrm{GeV}/c$ cut. Both distributions are shown on a logarithmic scale to highlight the wide dynamic range of particle momenta. The LHC distribution shows a broader and harder $p_T$ spectrum compared to RHIC. On the other hand, the $p_{T}$ distribution at $\sqrt{s} = 200\,\mathrm{GeV}$ is noticeably softer with a much steeper fall off as measured by the PHENIX Collaboration \cite{PHENIX:2011rvu}.}
\label{fig:pt_comparison}
\end{figure}

The key difference between the two sets is strikingly visible at the $p_T > 7\,\mathrm{GeV}/c$ cut, where the RHIC set has much better accuracy at 96.23\% compared to 91.33\% of the LHC set. This difference can be explained by the training of the model and the nature of the LHC set. The model has been trained on the LHC set at $\sqrt{s} = 13\,\mathrm{TeV}$, which means that the particles are much more energetic overall, with momentum ranges that are much higher than RHIC, especially at the transverse momentum range of $p_T > 7\,\mathrm{GeV}/c$. The performance of the model on different center of mass energies can be better understood by comparing the transverse momentum distributions in Figure~\ref{fig:pt_comparison}. A dataset with the condition of $p_T > 3\,\mathrm{GeV}/c$ includes a lot of high momentum particles, but relatively few in the $p_T > 7\,\mathrm{GeV}/c$ region, which contributes to the drop in accuracy for LHC. For any generated set, the transverse momentum distribution of particles peaks at lower values, corresponding to $3\,\mathrm{GeV}/c$ for the training set and $7\,\mathrm{GeV}/c$ for the test set.\\

\begin{figure}[H]
\centering
\begin{tabular}{cc}
\includegraphics[width=0.45\textwidth]{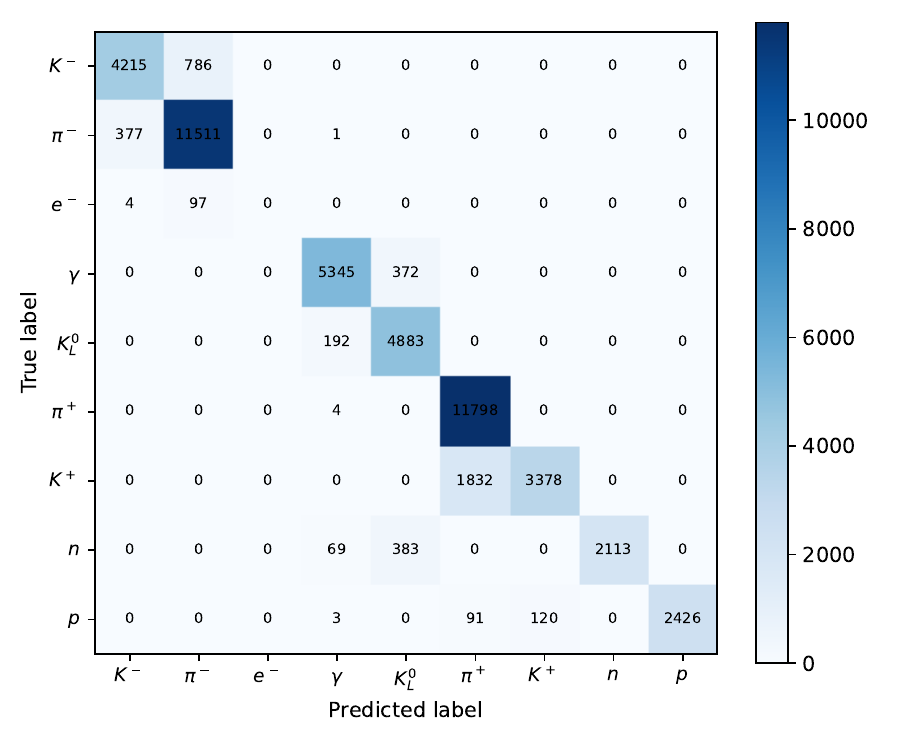} &
\includegraphics[width=0.45\textwidth]{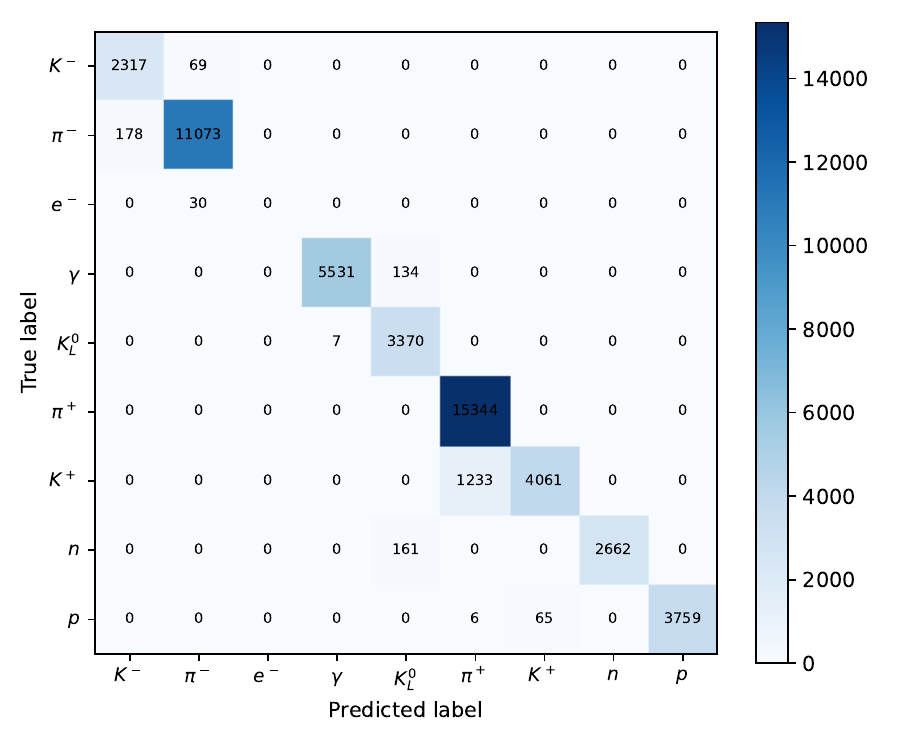} \\
\textbf{(a)} LHC $p_T > 7\,\mathrm{GeV}/c$ & \textbf{(b)} RHIC $p_T > 7\,\mathrm{GeV}/c$  \\
\end{tabular}
\caption{Comparison of confusion matrices between (a) LHC $\sqrt{s} = 13\,\mathrm{TeV}$ and (b) RHIC $\sqrt{s} = 200\,\mathrm{GeV}$ test sets with $p_T > 7\,\mathrm{GeV}/c$ cut. Despite similar overall accuracy, variations in particle-type accuracy and relative yields reveal the differences in production dynamics between the two energy regimes. }
\label{fig:confusion_comparison}
\end{figure}

\noindent
Figure~\ref{fig:confusion_comparison} further clarifies the classification process at high transverse momentum by comparing the confusion matrices of LHC and RHIC test sets with $p_T > 7\,\mathrm{GeV}/c$ cut. For both sets, there are expected drops in accuracy for most classes but the model retains decent separation ability in this quite complex regime. However, the two critical particle types, electrons and kaons,  highlight important underlying physics and model limitations.\\

\noindent
The model fails to correctly identify electrons at high $p_T$ for both test sets. This is primarily a physics limitation rather than a machine learning issue. In experimental settings, electrons are distinguished from pions using detector-level features such as the shower profile in electromagnetic calorimeters, the ratio of energy deposited in calorimeters to momentum ($E$/$p$) \cite{ALICE:2014sbx}, and the ionization energy loss per unit length ($\mathrm{d}E/\mathrm{d}x$) in time projection chambers \cite{YU201355}. Since the model only trained on kinematic features, it is realistically impossible to separate electrons from pions. Therefore, the model takes a conservative approach in trying to minimize the loss function by misidentifying electrons as the more abundant and kinematically similar class, namely the negatively charged pion ($\pi^-$). Since electrons represent a tiny fraction of the population for both sets, their misidentification has a negligible effect on the overall accuracy of the model.\\

\noindent
Another noteworthy trend occurs with kaons especially in the LHC test set. The positive kaon ($K^+$) has the second lowest accuracy across all particles at 64.83\%, with all misclassifications being positive pions ($\pi^+$). Even though these two particle types are significantly different in their rest mass and quark content, their kinematic signatures converge at such high transverse momentum and center-of-mass energy. In the $p_T > 7\,\mathrm{GeV}/c$ regime, the difference in rest mass becomes negligible compared to the available energy.\\

\noindent
Furthermore, a comparison between $K^+$ and $K^-$ in the confusion matrices reveals an asymmetric response in the model's behavior for these two particle types. Despite the fact that both are produced with nearly equal multiplicities in the LHC set, the model exhibits superior accuracy for $K^-$. This is clear in the LHC confusion matrix in Figure~\ref{fig:confusion_comparison} , where the $K^-$ identification accuracy is around 84.28\% compared with 64.83\% for ($K^+$). \\

\begin{figure}[H]
\centering
\begin{tabular}{cc}
\includegraphics[width=0.45\textwidth]{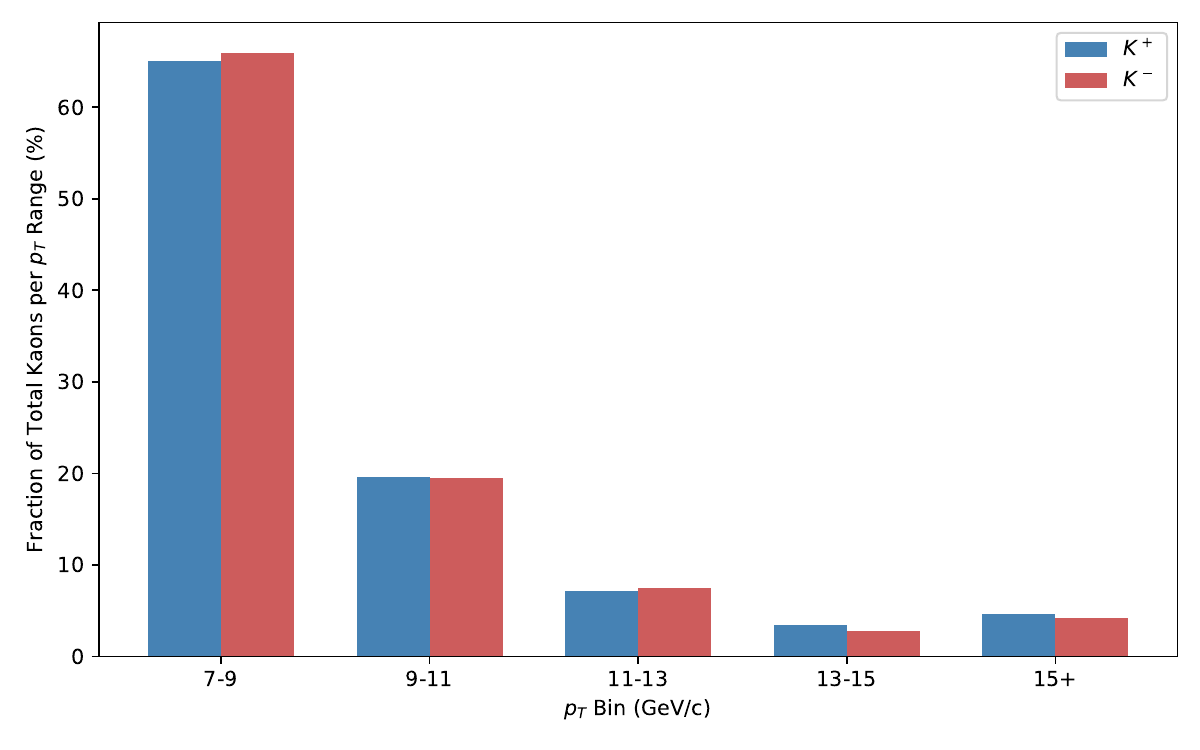} &
\includegraphics[width=0.45\textwidth]{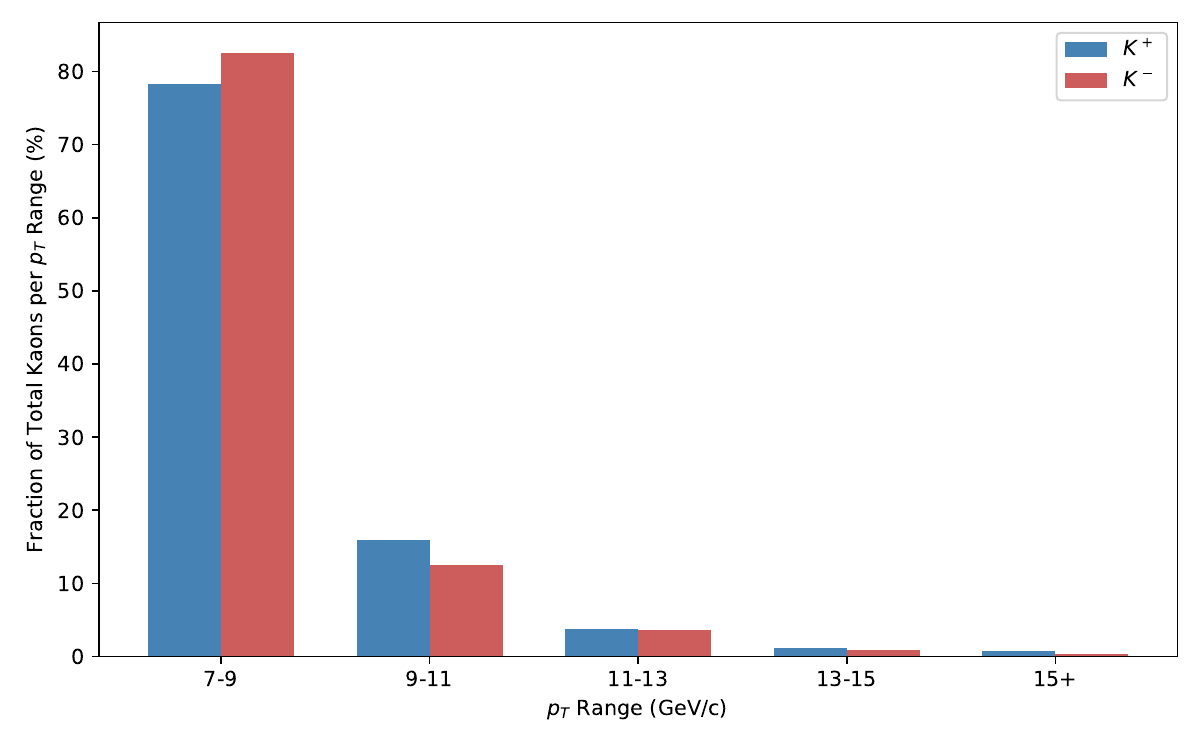} \\
\textbf{(a)} LHC kaon distribution  & \textbf{(b)} RHIC kaon distribution  \\
\end{tabular}
\caption{Fraction of total $K^+$ and $K^-$ for different $p_T$ bins for (a) LHC $\sqrt{s} = 13\,\mathrm{TeV}$ and (b) RHIC $\sqrt{s} = 200\,\mathrm{GeV}$ in the $p_T > 7\,\mathrm{GeV}/c$ test set. The counts are normalized separately for each species in both plots to illustrate their relative $p_T$ distributions rather than their exact numbers, since the $K^+$/$K^-$ varies significantly between the two energies.}
\label{fig:kaon_comparison}
\end{figure}

\noindent
This discrepancy in accuracy stems from the QCD nature of proton-proton collisions. $K^+$ mesons are produced more abundantly than $K^-$ mainly due to the leading particle effect, where hadrons sharing valence quarks with the colliding particle have a higher probability of being produced \cite{Braaten:2002yt}. The proton's valence quark structure ($uud$) makes it easier to produce a $K^+$ by combining one of the valence $u$-quarks with a $\Bar{s}$-quark from the quark sea. On the other hand, the production of a $K^-$ meson requires both quark $\Bar{u}$-quark and $s$-quark to be produced from the parton interactions. Moreover, this enhancement in $K^+$ production results in a broader $p_T$ distribution as shown in Figure~\ref{fig:kaon_comparison}. Negative kaons are mainly concentrated in the lowest possible $p_T$ bin. Across the vast majority of the higher $p_T$ bins, $K^+$ mesons dominate, which explains the significantly reduced accuracy of this particle type. \\

\noindent
Furthermore, comparing the RHIC and LHC $K^-$ distributions reveals the reason behind the model's superior performance on the $K^-$  mesons of the RHIC set. More than 80\% of negative kaons lie in the lowest $p_T$ bin compared to approximately 66\% for the LHC counterpart. The larger center-of-mass energy of LHC allows a larger fraction of $K^-$ to reach higher transverse momentum compared to RHIC.\\

\begin{figure}[H]
\centering
\begin{tabular}{cc}
\includegraphics[width=0.45\textwidth]{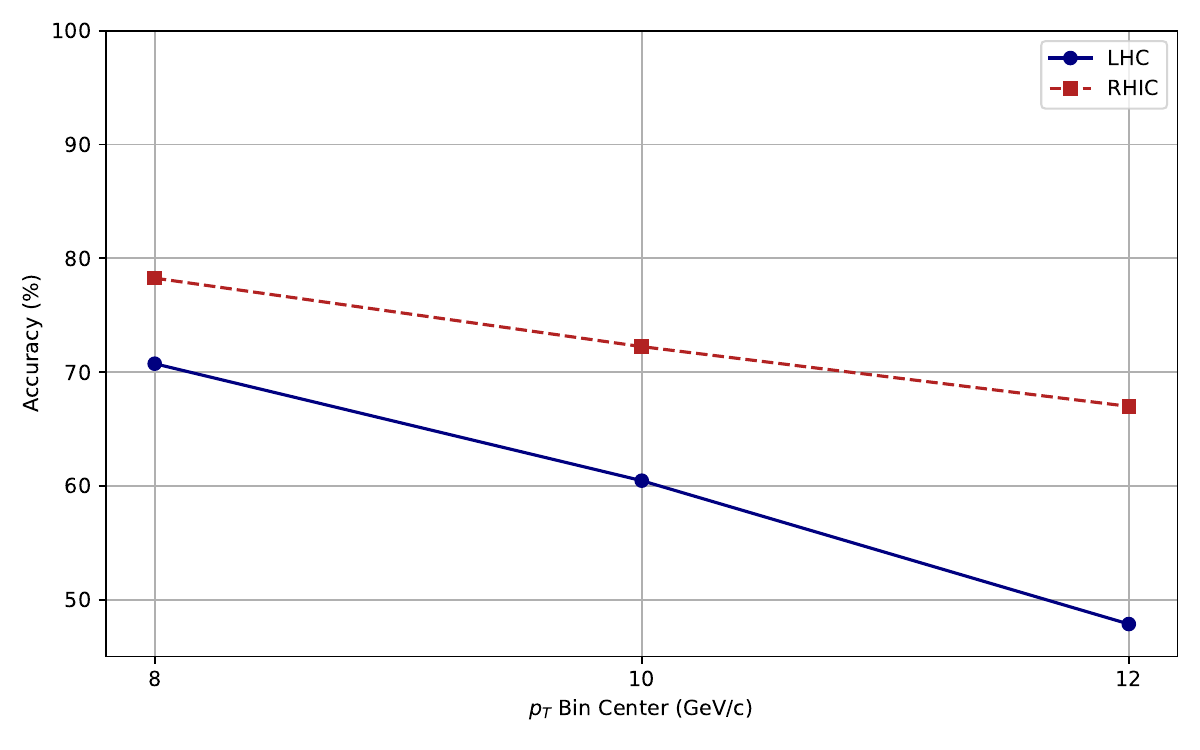} &
\includegraphics[width=0.45\textwidth]{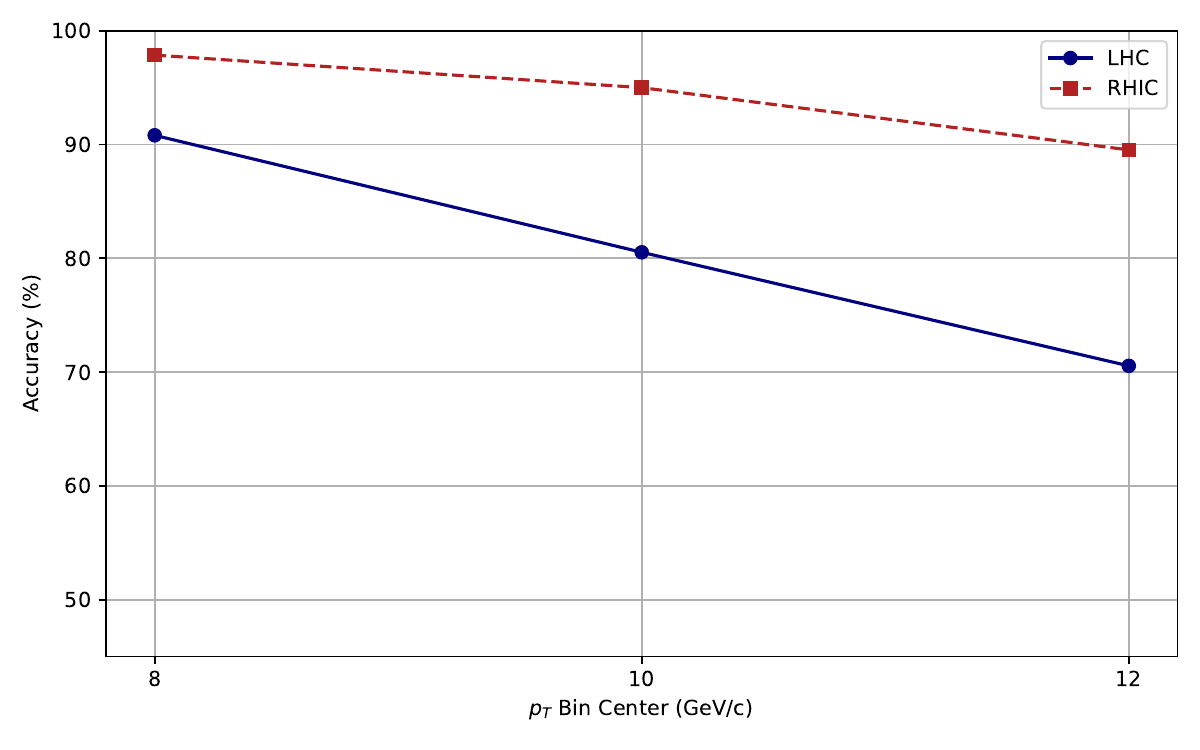} \\
\textbf{(a)} $K^+$ accuracy vs.\ $p_T$ bin centers  & \textbf{(b)}  $K^-$ accuracy vs.\ $p_T$ bin centers  \\
\end{tabular}
\caption{Per-class accuracy for $K^+$ and $K^-$ for different $p_T$ bins for both LHC and RHIC in the $p_T > 7\,\mathrm{GeV}/c$ test set. Accuracy is consistently higher for RHIC data within the same $p_T$ bins compared to LHC, which highlights fundamental differences in production dynamics between the two collision energies. A decrease in accuracy is observed for both $K^+$ and $K^-$ with increasing values of transverse momentum with the trend being more pronounced in the LHC case. }
\label{fig:kaon_accuracyvspt}
\end{figure}

\noindent
Figure~\ref{fig:kaon_accuracyvspt} shows per-class accuracy of $K^+$ and $K^-$  as a function of transverse momentum. The model performs significantly better on RHIC kaons compared to LHC across all the selected momentum bins, with the gap being wider at higher momenta. This result supports the interpretation that the model's classification ability is influenced by the underlying QCD production mechanisms, which differ meaningfully between center-of-mass energies. The lower available energy for RHIC restricts the kinematic reach of all particle species, reducing the kinematic overlap among particle types. These results suggest that the model's behavior is driven by physical features of high-energy collisions rather than memorization of training set kinematics.

\section{Conclusion}
\noindent
This study demonstrates the ability of a deep neural network trained exclusively on simulated LHC proton-proton collisions ($\sqrt{s} = 13\,\mathrm{TeV}$), to generalize effectively to RHIC data at a much lower center-of-mass energy. This generalization is achieved without any fine-tuning, weight adjustment or transfer learning. The model maintains an accuracy above 91\% at all momentum cuts, even in the complex $p_T > 7\,\mathrm{GeV}/c$ region. For the RHIC set under the same conditions, it achieves an accuracy of 96.23\%, which is quite notable as the model was not trained on any $\sqrt{s} = 200\,\mathrm{GeV}$ examples. \\

\noindent
The observed difference in the model's performance is consistent with expectations, given the higher kinematic reach of the LHC. Detailed analysis of per-class accuracy, particularly for kaons and electrons, reveals that the misidentification of particles is due to physical constraints related to production dynamics. The reduced classification accuracy for $K^+$ in high $p_T$ regions, especially at LHC energies, can be traced to the leading particle effect. This results in a broader and more energetic $p_T$ distribution relative to $K^-$, which in turn makes classification much more complex. Furthermore, complete misidentification of electrons at high $p_T$ arises from significant kinematic overlap with the much more common $\pi^+$ in the absence of discriminating detector-level features. The model's consistent performance across different center-of-mass energies suggests that it captures the physically meaningful features, rather than overfitting to the specific kinematics of the training energy scale. \\

\noindent
The model's robust performance from LHC regime to RHIC test conditions is far from trivial even within a controlled and simulated environment due to differences in the kinematic distributions and production dynamics, as revealed by the confusion matrices. This study represents a first step and a proof of concept for assessing how deep learning models  trained on higher center-of-mass energies behave when evaluated on lower collision energy data. The results warrant further investigation in real experimental settings to evaluate its practical viability. Applying this approach to real data naturally introduces multiple sources of uncertainty, including but not limited to detector resolution, reconstruction effects such as tracking inefficiencies, and domain shift between simulation and experiment. These challenges will need to be properly addressed in future studies.\\

\noindent
A key direction for future application is to enhance the current kinematic feature set with crucial detector-level observables, to address the model's difficulty in classifying certain particle types. On another front, future studies could explore more advanced network architectures which are better suited to deal with rich detector outputs. Prime candidates include convolutional neural networks which are ideal for energy deposition in calorimeters, ionization energy loss in TPC, and shower profiles. While CNNs represent a natural next step for analyzing spatial detector data, graph neural networks or transformer networks may offer more flexibility when dealing with complex topologies, sparse input features, and heterogeneous data types.

\noindent


\section{References}
\bibliography{bibliography.bib}{}
\bibliographystyle{unsrt}

\end{document}